\documentclass[%
 reprint,
 amsmath,amssymb,
 aps,
]{revtex4-2}

\usepackage{graphicx}
\usepackage{dcolumn}
\usepackage{bm}
\usepackage{xcolor}

\newcommand{\be}[1]{\begin{equation}\label{#1}}
\newcommand{\ee}{\end{equation}}
\newcommand{\ba}[1]{\begin{eqnarray}\label{#1}}
\newcommand{\ea}{\end{eqnarray}}
\newcommand{\rf}[1]{(\ref{#1})}
\newcommand{\nn}{\nonumber}

\begin{document}

\preprint{APS/123-QED}

\title{ Unification theory of instabilities of visco-diffusive swirling flows}

\author{Oleg N. Kirillov}
 \email{author for correspondence : oleg.kirillov@northumbria.ac.uk}
\affiliation{%
Northumbria University, Newcastle upon Tyne, NE1 8ST, United Kingdom
}%


\author{Innocent Mutabazi}
\affiliation{
Laboratoire Ondes et Milieux Complexes, UMR-6294 CNRS, Universit\'e Le Havre Normandie, Normandie Universit\'e, 53 Rue de Prony, 76058 Le Havre C\'edex, France
}%


\date{\today}

\begin{abstract}
 A universal theory of \textcolor{black}{linear} instabilities in swirling flows, occurring in both natural settings and industrial applications, is formulated. The theory encompasses a wide range of open and confined flows, including spiral isothermal flows and baroclinic flows driven by radial temperature gradients and natural gravity in rotating fluids. By employing short-wavelength local analysis, the theory generalizes previous findings from numerical simulations and linear stability analyses of specific swirling flows, such as spiral Couette flow, spiral Poiseuille flow, and baroclinic Couette flow. A general criterion, extending and unifying existing criteria for instability to both centrifugal and shear-driven perturbations in swirling flows is derived, taking into account viscosity and thermal \textcolor{black}{diffusion, and guiding} experimental and numerical investigations in the otherwise inaccessible parameter regimes.

\end{abstract}

\maketitle
\textit{Swirling flows} induced by the combination of rotation and shear in orthogonal directions are ubiquitous in various natural phenomena, such as tornadoes and tropical cyclones \cite{E2018,Emmanuel}, meandering rivers \cite{PA2019}, vortex rings with swirl \cite{LH1993}, geophysical and astrophysical flows \cite{Knobloch,Busse,Lopez2013},  and recently discovered magnetic tornadoes in the solar atmosphere \cite{Wedemeyer,Rogava}. These flows also occur in trailing vortices of aircraft wingtips \cite{Feys,Leibovich,ES1978,BG2013,E1984} and in branching junctions of everyday piping systems and physiological flows \cite{A2016}, where identifying instabilities that lead to vortex breakdown is of paramount importance \cite{LNOD2001}. Swirling flows are present in industrial processes, such as filtration or purification of wastewater \cite{Ollis1991}, isotope separation through centrifugation \cite{Lamarsh}, and  oil-drilling systems \cite{MacAndrew}. They are also characteristic of convective flows with rotation \cite{Lappa}, associated with cooling or lubrication of rotating machinery \cite{Kreith,Lee1989,Fenot2011,SLW2022}, crystal growth \cite{Springer}, and solidification of metals \cite{vives}.

From a hydrodynamic perspective, the base state of a swirling flow has azimuthal and axial velocity components in either open or confined geometries. The open flow configuration is typical of swirling jets in natural phenomena, while the confined one is more common in engineering. A convenient setup to study swirling flows, both theoretically and experimentally, confines the fluid in a cylindrical annulus with differentially rotating cylinders, creating the classical circular Couette flow. \textcolor{black}{In this setup, axial velocity can be driven by an external pressure gradient, as in Spiral Poiseuille flow (SPF), which combines annular Poiseuille flow with circular Couette flow \cite{TJ1981, CP2004, Messeguer1, MM2005, CM2005, H2008a}. Alternatively, axial flow can be induced by sliding the inner cylinder, producing Spiral Couette flow (SCF), which combines sliding Couette flow with circular Couette flow \cite{L1964, Ng, Ali2, MM2000}, or by imposing a radial temperature gradient, which generates baroclinic vorticity, resulting in baroclinic Couette flow (BCF) \cite{Snyder, Ali1, Lepiller, Yoshikawa, Guillerm, Kang2015, Kang2023}.} While SPF and SCF are stationary solutions to the Navier-Stokes equations, BCF is a stationary solution to the Navier-Stokes equations coupled with the energy equation via buoyancy and transport terms.

In the absence of axial flow and a radial temperature gradient, a differentially rotating azimuthal flow of inviscid, incompressible fluid becomes centrifugally unstable if it meets the Rayleigh criterion, which states that angular momentum must decrease outward for instability to occur \cite{EY1995}. Including viscosity only slightly reduces the instability domain defined by this result \cite{EY1995}. 
\textcolor{black}{However, the presence of a radial temperature gradient and thermal diffusion can destabilize Rayleigh-stable flows, such as quasi-Keplerian flows \cite{Kirillov2017, MMY2021}, through the Goldreich-Schubert-Fricke (GSF) instability—a visco-diffusive extension of Rayleigh's centrifugal instability that applies to non-isothermal, differentially rotating azimuthal flows \cite{M2013,JT2023}. The GSF instability may significantly influence angular momentum transport in stellar radiation zones and astrophysical discs, driving turbulence and stirring solid materials in regions of protoplanetary discs that are stable against magnetorotational instability \cite{Petersen2007,Klahr2014,Held2018,Balbus,Ji2022,S2024,Kirillov2013,SK2015}.
Another destabilization mechanism in non-isothermal, differentially rotating azimuthal flows is the oscillatory visco-diffusive McIntyre instability \cite{M1970,LK2021}.}

In the absence of rotation, axial flows in a cylindrical annulus are subject to wall-driven shear instabilities. Annular Poiseuille flow becomes linearly unstable above a critical axial Reynolds number for all $0 < \eta \le 1$, where $\eta$ is the ratio of the inner to outer cylinder radii \cite{MFN2008,H2008}. For  sliding Couette flow, the critical  Reynolds number is finite for $0 < \eta < 0.1415$ and infinite otherwise, with the unstable mode being axisymmetric \cite{Deguchi}. Baroclinic convection in a vertical annulus with a horizontal temperature gradient becomes linearly unstable to predominantly axisymmetric perturbations above a finite axial Grashof number for $0.01 \le \eta \le 0.99$ when the Prandtl number, the ratio of viscosity to thermal diffusivity, is between 1 and 30 \cite{Bahloul2000, Lepiller2007, WC2022}.

In contrast, the swirling flows  can become unstable due to both centrifugally-driven and shear-driven perturbations. The competition between these destabilizing mechanisms leads to a rich variety of flow patterns and bifurcations, as documented in the scattered experimental, theoretical, and numerical studies \cite{TJ1981, CP2004,  Messeguer1, MM2005,CM2005,H2008a, L1964, Ng, Ali2, MM2000,Snyder, Ali1, Lepiller,Yoshikawa,Guillerm,Kang2015,Kang2023}.

\textcolor{black}{In isothermal swirling flows, an intriguing phenomenon has been observed: the critical Reynolds number of the inner cylinder in Couette flow shows a sudden drop in the Rayleigh-stable regime as the axial Reynolds number increases. Surprisingly, this behavior consistently appears in both spiral Poiseuille \cite{Messeguer1} and spiral Couette \cite{MM2000} flows, despite differences in their shear profiles. The underlying mathematical reason for this phenomenon involves a \textit{pleat} and \textit{folds} in the critical surface that defines the neutral stability boundary within the space of two azimuthal Reynolds numbers and the axial Reynolds number—determined by the sliding speed of the inner cylinder for SCF and the mean axial flow velocity for SPF. This characteristic geometry aligns with the \textit{cusp catastrophe} model \cite{BG1992} and reflects the interplay between centrifugal and shear instabilities in swirling flows. Detecting the pleat and the folds in \cite{MM2000, Messeguer1} required developing a \textit{specifically tailored numerical scheme}.}

The only general \textit{analytical} instability criterion in this context is available for isothermal inviscid incompressible swirling flows. It is the extension of Rayleigh's criterion, derived via a global modal approach by Ludwieg \cite{L1964} for SCF in the narrow gap limit and by Leibovich and Stewartson \cite{Leibovich} for swirling jets. Eckhoff \cite{ES1978,E1984} arrived at the same criterion using the local geometrical optics approach \cite{LH1993}, deriving instability conditions by tracing a localized wave packet along the streamlines of the helical base flow. Leblanc and Le Duc \cite{LL2005} established a formal equivalence between the local and global approaches to the Ludwieg-Eckhoff-Leibovich-Stewartson (LELS) criterion  in the limit of large wavenumbers. However, the LELS criterion, limited to inviscid flows, cannot describe the folded neutral stability surface of SCF and SPF, nor can it be applied to non-isothermal swirling flows. A theory unifying these phenomena has yet to be formulated.

\textit{This Letter} presents a general theory for investigating instabilities of both viscous isothermal and visco-diffusive non-isothermal swirling flows using the local geometrical optics stability analysis pioneered in the hydrodynamics of inviscid flows \cite{ES1978, LH1993, FV1991, IK2017} and recently extended to visco-diffusive flows \cite{M1986, Kirillov2017,Kir17prsa,LK2021, KSF2014}. We apply this theory to the helical stationary solutions of the Navier-Stokes equations coupled with the energy equation in the Boussinesq approximation, which, depending on the boundary conditions, can represent SCF, SPF, or BCF.

We establish that the neutral stability curves in the plane of the azimuthal and axial Reynolds numbers form families with a consistent \textit{envelope}, regardless of whether they are parameterized by the azimuthal or axial wavenumber. Since the axial wavenumber can take arbitrary real value, this envelope defines the boundary of the union of individual instability domains for specific wavenumbers, providing a comprehensive instability criterion that includes the Rayleigh, GSF, and LELS criteria as special cases.

We find that the envelope, and thus the instability domain, splits during the transition from Rayleigh-unstable to Rayleigh-stable flows. For Rayleigh-stable flows, as the azimuthal Reynolds number approaches infinity, an asymptotic line to the envelope represents the inviscid LELS criterion for the isothermal flows and yields an analytical expression for the new instability criterion for the non-isothermal flows, uniting the LELS and the GSF criteria. In the isothermal case, we find a compact closed-form expression for the envelope, generalizing the inviscid LELS criterion to viscous swirling flows and covering the full range of azimuthal Reynolds numbers from zero to infinity. This extension broadens the criterion's applicability to previously inaccessible, practically important situations and offers an exact analytical expression for the folded neutral stability surface, \textit{universal for all isothermal swirling flows}.

\textcolor{black}{We consider an incompressible Newtonian fluid with constant reference density $\rho$ as well as constant thermal expansion coefficient $\alpha$, kinematic viscosity $\nu$, and thermal diffusivity $\kappa$.} This fluid is confined within an infinitely long cylindrical annulus with a gap width $d = R_2 - R_1$, where $R_1$ is the radius of the inner cylinder at temperature $T_1$, rotating with angular velocity $\Omega_1$, and $R_2$ is the radius of the outer cylinder at temperature $T_2 = T_1 - \Delta T$, rotating with angular velocity $\Omega_2$. We denote $\eta=R_1/R_2$ and $\mu=\Omega_2/\Omega_1$. The system is subjected to a uniform gravity field with acceleration $g$ along the $Z$-axis of the cylindrical coordinates $(R, \varphi, Z)$, which aligns with the common rotation axis of the cylinders.

\textcolor{black}{We assume the base flow to be helical, characterized by azimuthal and axial velocity components only. Setting the inner cylinder velocity, \( V_0 = \Omega_1 R_1 \), as the velocity scale,} $d$ as the length scale, $d/V_0$ as the time scale, and $\rho V_0^2$ as the pressure scale, and  applying the Boussinesq-Oberbeck approximation—which assumes all fluid properties are constant except for the density, which varies linearly with temperature in the driving forces—we write the dimensionless governing equations:
\ba{nle}
&\nabla \cdot \boldsymbol{u}=0,&\nn\\
&\frac{d \boldsymbol{u}}{d t}+\nabla p-\frac{1}{Re}\nabla^2\boldsymbol{u}+
\left(\gamma \frac{v^2}{r}\boldsymbol{e}_r-Ri\boldsymbol{e}_z\right)\theta=0,&\nn\\
&\frac{d \theta}{d t}-\frac{1}{RePr}\nabla^2\theta=0,&
\ea
where $p$ and $\boldsymbol{u}=(u,v,w)$ are the dimensionless pressure and velocity field, respectively, $\theta = \frac{T-T_2}{\Delta T}$ is the temperature deviation,
$\frac{d}{dt} = \frac{\partial}{\partial t} +{\bm u} \cdot \nabla$, $r=\frac{R}{d}$, and $z=\frac{Z}{d}$.

The dimensionless control parameters in Eq.~\rf{nle} are
\ba{dlp}
\quad Re=\frac{V_0 d}{\nu},
\quad Pr=\frac{\nu}{\kappa},
\textcolor{black}{\quad Ri=\frac{Gr}{Re^2},
\quad Gr=\frac{W_T d}{\nu},}
\ea
where $Re$ is the Reynolds number associated with the rotation of the inner cylinder, $Pr$ is the Prandtl number, $Ri$ is the Richardson number \textcolor{black}{associated with the Archimedean buoyancy term, and $Gr$ is the Grashof number, which characterizes the strength of the baroclinic flow. The Grashof number} is defined using the characteristic thermal velocity $W_T=\frac{\gamma g d^2}{\nu}$ and $\gamma=\alpha \Delta T$, with $\gamma>0$ for outward heating $(T_1 > T_2)$.

\textcolor{black}{In a cylindrical annulus with an infinite aspect ratio, \(\left(\Gamma = \frac{L}{d} \rightarrow \infty\right)\), where \(L\) is the length of the annulus, the base flow state remains stationary and invariant along the axial direction.} Its two-component dimensionless velocity field is $\boldsymbol{u}_B(r)=(0,V(r),S^{-1}W(r))$, where $S=\frac{V_0}{W_0}$ is the swirl parameter, \textcolor{black}{and \(W_0\) represents the characteristic axial velocity defined individually for each base flow \cite{Ali1,Ali2}.} The temperature $\theta_B(r)=\Theta(r)$ depends solely on the radial coordinate $r$, while the pressure is expressed as $p_B(r,z)=P_1(r)+z P_2$ \cite{Ali1,Ali2}.

To produce the base state of BCF, we set $W_0=W_T$ and impose the boundary conditions: $V(r_1)=1$, $V(r_2)=\mu/\eta$, $W(r_{1,2})=0$, and $\Theta(r_1)=1$, $\Theta(r_2)=0$, where $r_1=\frac{\eta}{1-\eta}$ and  $r_2=\frac{1}{1-\eta}$. The azimuthal velocity and the temperature profiles of BCF are given by:
\be{VT}
    V(r)=Ar+\frac{B}{r},\quad \Theta(r)=\frac{\ln\left[\left(1-\eta\right)r\right]}{\ln\eta}
\ee
with
$A= \frac{\mu-\eta^2}{\eta\left(1+\eta \right)}$ and $B= \frac{\eta}{1+\eta}\frac{1-\mu}{\left(1-\eta \right)^2}$. In addition to the two integration constants fixed by the boundary conditions, the axial profile  $W(r)$ of BCF depends on the gradient  $P_2$, which is determined by the condition of zero axial mass flux $\int_{r_1}^{r_2}rW(r)dr=0$, and reads \cite{Ali1}:
\ba{W}
&W(r)= C\left[(r_2^2{-}r_1^2)\frac{\ln(r/r_2)}{\ln \eta} {+} r^2 - r_2^2\right]-(r^2 {-} r_1^2)\frac{\ln(r/r_2)}{4\ln\eta},&\nn\\
&C = \frac{1}{16}\frac{\left(1-3\eta^2\right)\left(1-\eta^2\right)-4\eta^4\ln\eta}{\left(1-\eta^2\right)^2+\left(1-\eta^4\right)\ln\eta}.&
\ea
\textcolor{black}{Introducing the axial Reynolds number, \(Re_z = \frac{W_0 d}{\nu}\), with \(W_0 = W_T\), gives \(Re_z = Gr\) and \(S = \frac{Re}{Gr}\).}
\begin{figure*}[t]
\includegraphics[width=.9\textwidth]{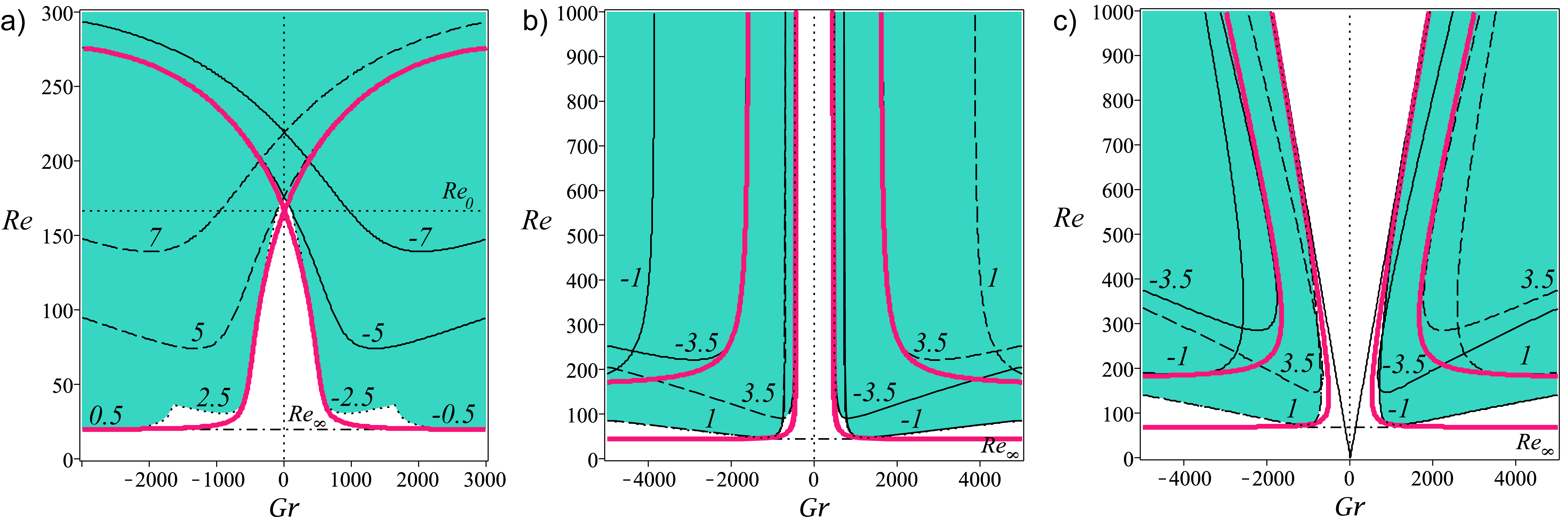}
\caption{Neutral stability curves \(\rf{a0}\) parameterized by \(k_z\), with their envelope (red, thick) for the non-isothermal BCF. Parameters are \(Pr = 5.5\), \(\gamma = 0.0004\), \(k_r = 1.7\pi\), and \(\eta = 0.8\) (\textcolor{black}{selected to facilitate comparison with numerical results from previous studies \cite{Lepiller,Yoshikawa,Guillerm,Kang2015,Kang2023}}) for the three different $\mu$: (a) Rayleigh unstable, \(\mu = 0\); (b) modified Rayleigh line, \(\mu = \mu_R \approx 0.63935\) (from \(\rf{mur}\)); and (c) Rayleigh stable, \(\mu = 0.8\). The envelope at \(Gr = 0\) in (a) gives \(Re_0 \approx 166.8\) from \(\rf{re0}\). \textcolor{black}{Green shaded regions are the unions of instability domains for specific \(k_z\).} Solid oblique straight lines in (c) represent the new unified criterion \(\rf{elsscf1}\). All computations for the BCF are performed at the mean geometric radius \(r = \sqrt{r_1 r_2}\).
\label{RayleighRBF}}
\end{figure*}

To test the stability of the base flow \rf{VT}, \rf{W}, we consider its small three-dimensional perturbations $\left(\boldsymbol{u}', p',\theta'\right)$ and linearize the nonlinear equations \rf{nle} about the base BCF state.
Introducing a small parameter $0<\epsilon \ll 1$, we
represent the perturbations as asymptotic expansions:
\ba{pert}
&\boldsymbol{u}^{\prime} =\left(\boldsymbol{u}^{(0)}(\boldsymbol{x}, t){+}\epsilon \boldsymbol{u}^{(1)}(\boldsymbol{x}, t)\right){e}^{\frac{{i} \Phi(x, t)}{ \epsilon}}+\epsilon \boldsymbol{u}^{(r)}(\boldsymbol{x}, t,\epsilon),&\nn\\
&\theta^{\prime} =\left(\theta^{(0)}(\boldsymbol{x}, t){+}\epsilon \theta^{(1)}(\boldsymbol{x}, t)\right){e}^{\frac{{i} \Phi(x, t)}{ \epsilon}}+\epsilon \theta^{(r)}(\boldsymbol{x}, t,\epsilon),&\nn\\
&p^{\prime} =\left(p^{(0)}(\boldsymbol{x}, t){+}\epsilon p^{(1)}(\boldsymbol{x}, t)\right){e}^{\frac{{i} \Phi(x, t)}{ \epsilon}}+\epsilon p^{(r)}(\boldsymbol{x}, t,\epsilon),&
\ea
where ${i}{=}\sqrt{-1}$ and $\bm{u}^{(r)}$, $p^{(r)}$, $\theta^{(r)}$ are uniformly bounded in $\epsilon$ \cite{ES1978,FV1991,LH1993,KSF2014}. 

\textcolor{black}{Maslov \citep{M1986} observed that high-frequency oscillations of the form $\exp\left( i\epsilon^{-1}\Phi(\boldsymbol{x}, t)\right)$ quickly decay due to viscosity, unless a quadratic dependence of viscosity on the small parameter \(\epsilon\) is assumed. Therefore, following Maslov, we assume \(Re \sim \epsilon^{-2}\) in the linearized equations (see also \cite{Kirillov2013,KSF2014,SK2015,Kir17prsa,Kirillov2017,LK2021}). Substituting \rf{pert} into the linearized equations and collecting terms with matching powers of \(\epsilon\), we derive a hierarchical system of equations to determine the coefficients \(\boldsymbol{u}^{(0)}\), \(\theta^{(0)}\), \(p^{(0)}\), \(\boldsymbol{u}^{(1)}\), and so on, in the expansions \rf{pert}.}

At orders \(\epsilon^{-1}\) and \(\epsilon^{0}\), the pressure can be eliminated.
This simplification yields equations in the stationary frame for the amplitudes \(\boldsymbol{u}^{(0)}\), \(\theta^{(0)}\) of the localized wave packets moving along the streamlines of the base flow:
\ba{eq19}
&\frac{d {\bm u}^{(0)}}{dt}+\frac{|{\bm k}|^2}{Re}{\bm u}^{(0)}
=-\left(\mathcal{I}-\frac{{\bm k}{\bm k}^T}{|{\bm k}|^2}\right)\left(\gamma\frac{V^2}{r}{\bm e}_r-Ri{\bm e}_z\right)\theta^{(0)}&\nn\\
&-\left(\mathcal{I}-2 \frac{{\bm k}{\bm k}^T}{|{\bm k}|^2}\right) \mathcal{U}{\bm u}^{(0)}-2\gamma \Theta\Omega \left(\mathcal{I}-\frac{{\bm k}{\bm k}^T}{|{\bm k}|^2}\right){\bm e}_r{\bm e}_{\varphi}^T{\bm u}^{(0)},&\nn\\
&\frac{d \theta^{(0)}}{dt}+\frac{|{\bm k}|^2}{RePr}\theta^{(0)}=-\left(\nabla\Theta\right)^T{\bm u}^{(0)},&
\ea
where the wavevector $\bm{k}{=}\frac{\bm{\nabla}{\Phi}}{\epsilon}$ satisfies the equation
\be{eik}
 \frac{d\bm{k}}{dt}=-\mathcal{U}^T{\bm k}, \quad \mathcal{U}=\left(\begin{array}{ccc}
0 & -\Omega & 0 \\
\Omega(1+2Ro) & 0 & 0 \\
S^{-1}DW & 0 & 0 \\
\end{array}\right)
\ee
with $\mathcal{I}$ the identity matrix, ${\bm k}{\cdot} {\bm u}^{(0)}{=}0$, $\frac{d}{dt} = \frac{\partial}{\partial t} +{\bm u}_B \cdot \nabla$, $D=\frac{d}{dr}$, \textcolor{black}{$\Omega(r)=\frac{V(r)}{r}$}, and the Rossby number $Ro=\frac{rD\Omega}{2\Omega}$.

In the frame of the wave packet, rotating about the vertical axis with the angular velocity \(\Omega(r)\), the wavevector ${\bm k}=(k_r,k_{\varphi},k_z)$ is time-independent and the amplitude equations \rf{eq19} are autonomous, provided that \cite{ES1978,Leibovich,Emmanuel}
\be{ec}
k_\varphi=-\overline{DW} k_z, \quad {\rm with} \quad \overline{DW}=\frac{DW}{2\Omega RoS}.
\ee
Thus, we seek the most unstable and exponentially growing 3D perturbations in swirling flows. These perturbations display helical symmetry \cite{Emmanuel}, remaining invariant along circular helices with a pitch of $2\pi r \overline{DW}$ \cite{ES1978,Leibovich,BG2013}. \textcolor{black}{Note that the equations \rf{eq19}-\rf{ec} apply for each helical streamline of the base flow.}

\textcolor{black}{In this study, we focus on the development of normal modes in the flow, disregarding transient modes, although the latter can be dangerous in open flows such as SPF \cite{HP2006, H2008a}.}
Assuming $\bm{u}^{(0)}, \theta^{(0)}\sim e^{s t+i m\varphi+ik_z z}$ in \rf{eq19},
where $s=\sigma + i \omega$ \textcolor{black}{($\sigma,\omega \in \mathbb{R}$)} is the complex growth rate, and $m=k_\varphi r$ and $k_z$ are the integer azimuthal and real axial wavenumbers of the perturbations,
and taking into account \rf{ec}, we obtain the dispersion relation $\det(\mathcal{H}-\lambda\mathcal{I})=0$, where $|\bm{k}|^2=k_r^2+k_z^2\left[1+\overline{DW}^2\right]$, $\lambda=\sigma+i\left(\omega+m\Omega+k_zW/S \right)$, and
\begin{widetext}
\ba{mahc}
&\mathcal{H}= \left(
          \begin{array}{ccc}
            -\frac{ DW}{ SRo}\frac{k_r k_z}{|{\bm k}|^2}-\frac{|{\bm k}|^2}{Re} & 2\Omega(1-\gamma \Theta)\left(1-\frac{k_r^2}{|{\bm k}|^2}\right) & -\left(r \gamma \Omega^2
            \left(1-\frac{k_r^2}{|{\bm k}|^2} \right) + Ri\frac{k_r k_z}{|{\bm k}|^2}\right)  \\
            -2\Omega\left(Ro+\frac{k_r^2+k_z^2}{|{\bm k}|^2} \right) & (1-\gamma\Theta)\frac{ DW}{ S Ro}\frac{k_r k_z}{|{\bm k}|^2} -\frac{|{\bm k}|^2}{Re}& -\left(r  \gamma \Omega^2   - Ri\frac{k_z}{k_r}\right)\frac{1 }{2\Omega} \frac{ DW}{S Ro}\frac{k_r k_z}{|{\bm k}|^2}  \\
            -D\Theta  & 0 & \frac{|{\bm k}|^2}{Re} \frac{Pr-1}{Pr}-\frac{|{\bm k}|^2}{Re} \\
          \end{array}
        \right).&
\ea
\end{widetext}
The dispersion relation $\lambda^3+a_2\lambda^2+a_1\lambda+a_0=0$, defined by  \rf{mahc}, has real coefficients. Therefore, the stability conditions of the base flow ($\sigma < 0$) are defined by the \textcolor{black}{Li\'enard-Chipart} stability criterion: $a_0 >0$, $a_2>0$, $a_0(a_1a_2-a_0)>0$  \cite{Kirillov2017}. \textcolor{black}{The threshold for the onset of oscillatory instability, relevant to the visco-diffusive McIntyre instability \cite{M1970}, is obtained from the last inequality of this criterion \cite{Kirillov2017, LK2021}, with a detailed analysis left for future work.} \textcolor{black}{The condition}
\begin{widetext}
\ba{a0}
&a_0 =  \frac{D\Theta DW}{SRo} \frac{k_z^2}{|{\bm k}|^2}  Ri  (1-\gamma \Theta )-\frac{k_r k_z}{|{\bm k}|^2}\frac{|{\bm k}|^2}{Re}RiD\Theta
+\frac{|{\bm k}|^2\Omega^2}{Re Pr}\left(1 - \frac{k_r^2}{|{\bm k}|^2}\right)
\left\{4\left[Ro + \frac{k_r^2 + k_z^2}{|{\bm k}|^2}\right](1-\gamma \Theta ) -  \gamma  r D\Theta Pr \right\}&\nn\\
&+\frac{|{\bm k}|^2}{ Re Pr}\left( \frac{|{\bm k}|^2}{ Re}-\frac{k_rk_z}{|{\bm k}|^2}\frac{DW}{SRo}(1-\gamma\Theta) \right)
\left(\frac{|{\bm k}|^2}{ Re} + \frac{k_rk_z}{|{\bm k}|^2}\frac{DW}{SRo}\right)=0,&
\ea
\end{widetext}
\textcolor{black}{yields $\lambda=0$, corresponding to the modes with $\sigma=0$ and the frequency $\omega=-m\Omega-k_zW/S$.} The marginal modes are inclined on the cylindrical surface of radius $r$ and have azimuthal and axial phase velocities $c_\varphi{=}-\Omega r$ and $c_z{=}-k_zW/S$. With $Ri=\frac{1}{ReS}$, the limit $S\rightarrow \infty$ of \rf{a0} retrieves the  critical value of $Re$ for stationary modes of the thermal convection induced by centrifugal buoyancy in the differentially rotating cylindrical annulus \cite{Kirillov2017}.

In general, \(\rf{a0}\) defines a family of marginal stability curves in the \((Gr, Re)\)-plane, parameterized by the axial wavenumber \( k_z \) (or equivalently \( m \) via \(\rf{ec}\)).

 \begin{figure*}[t]
\includegraphics[width=.95\textwidth]{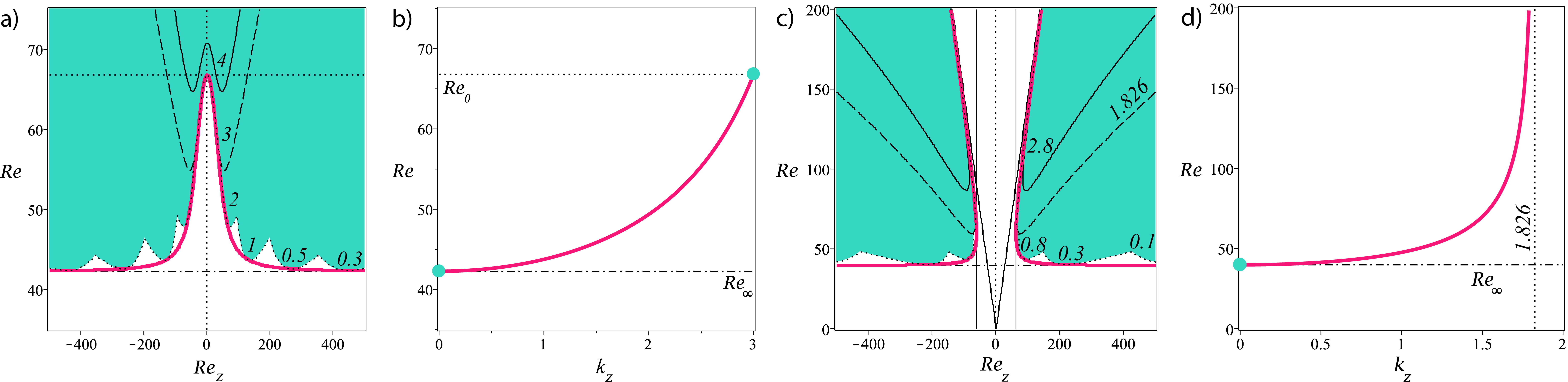}
\caption{ (a, c) \textcolor{black}{Green shaded area represents the union of individual instability domains defined by the neutral stability curves \(\rf{a0}\) in the \((Re_z, Re)\)-plane for (a) \(k_z=0.3, 0.5, 1, 2, 3, 4\) and (c) \(k_z=0.1, 0.3, 0.8, 1.826, 2.8\).} The thick red curves show their envelope \(\rf{masterisc}\) for the isothermal enclosed SCF \rf{escf} with \(\eta = 0.4\) and \(k_r = 3\sqrt{2}\). Panels (a, b) represent \(\mu = 0\), and panels (c, d) represent \(\mu = 0.5\). The black dashed curves correspond to the terminal \(k_z = \frac{\sqrt{2}}{2} k_r = 3\) in (a) and the terminal \(k_z = \frac{\sqrt{-2Ro}}{2} k_r \approx 1.826\) in (c). \textcolor{black}{The black solid curves in (a,c) show the boundaries of the individual instability domains corresponding to $k_z$ exceeding the terminal value.} The oblique black solid lines in (c) indicate the inviscid LELS criterion \rf{lels}. Vertical solid lines in (c) show $\pm Re_z^{\min}$ \(\rf{rezm}\). (b,d) Variation of \(k_z\) from $0$ to the terminal value according to \(\rf{kze}\) as \(Re\) increases from \(Re_{\infty}\) (dot-dashed line) to (b) \(Re_0\) (dotted line) or (d) to infinity. All computations for the SCF are performed at the mean geometric radius \(r = \sqrt{r_1 r_2}\).
\label{Figure2}}
\end{figure*}

\textit{Our first key finding}, overlooked in previous works, is that the individual marginal stability curves of the BCF have an envelope determined by computing the discriminant of \(\rf{a0}\), treated as a polynomial in \(k_z\) or \(m\) \cite{BG1992,Hartman}.
This envelope, independent of the chosen parameterization, separates the domain of unstable modes from the stability zone. It consists of two distinct symmetric curves that can intersect at $Gr = 0$ and $Re=Re_0$, where \be{re0}
    Re_0=\frac{3\sqrt{3}k_r^2}{2\Omega}\frac{1}{\sqrt{rD\Theta\gamma Pr -4\left(1+Ro\right)\left(1-\gamma \Theta \right)}},
\ee
as shown in Fig.~\ref{RayleighRBF}(a). \textcolor{black}{Expressed in terms of the Taylor number, defined as $ Ta = \frac{2Re \Omega}{3\sqrt{3}k_r^2} $, equation \eqref{re0} precisely matches the result obtained in \cite{Kirillov2017} for circular Couette flow with a radial temperature gradient.} 

\textcolor{black}{The condition \( a_0 < 0 \) defines the instability regions for each \( k_z \) (or \( m \)), with the union of these regions for \( k_z = \pm 0.5\), \( \pm 2.5\), \( \pm 5\), and \( \pm 7\) represented as the green shaded area in Fig.~\ref{RayleighRBF}(a). The boundaries of the individual instability domains, which meet the envelope at \( Re > Re_0 \) and thus lie entirely within the green shaded area, are shown as dashed lines for \( k_z = 5 \) and \( 7 \), and as solid lines for \( k_z = -5 \) and \( -7 \) in Fig.~\ref{RayleighRBF}(a). This notation emphasizes that the absolute values of these axial wavenumbers exceed the terminal wavenumber, $k_z=\frac{\sqrt{2}}{2}k_r$, where they contact the envelope at \( Re = Re_0 \). In Fig.~\ref{RayleighRBF}(b,c), the dashed and solid curves mark the boundaries of the individual instability domains (which union is shown as a shaded green area) for \( k_z = 1\) and \( 3.5\), and \( k_z = -1\) and \( -3.5\), respectively.}

The requirement for the radicand in  \rf{re0} to be positive defines the modified Rayleigh line for non-isothermal flows:
\be{mur}
 Ro_R = -1+\frac{rD\Theta\gamma Pr }{4\left(1{-}\gamma \Theta\right)} = -1+\frac{1}{4} \gamma Pr\, rD\Theta +O(\gamma^2).
\ee

While intersecting for $Ro<Ro_R$, the envelope branches have vertical asymptotes at the modified Rayleigh line \rf{mur}, delimiting the zone where no instability modes can be obtained, Fig.~\ref{RayleighRBF}(b). For $Ro>Ro_R$ the asymptotes are inclined, placing the instability domains within the half-planes defined by the asymptotes to the external branch, shown by oblique lines in Fig.~\ref{RayleighRBF}(c):
\be{elsscf1}
\frac{N_{\Omega}^2}{\Omega^2}(1-\gamma\Theta)+Pr\frac{N^2}{\Omega^2} <
\frac{\left( \overline{DW}(2-\gamma\Theta)-\frac{Pr  D\Theta}{2k_r^2\Omega  S}\right)^2}{1 + \overline{DW}^2}.
\ee
Here $N_{\Omega}^2=4(1+Ro)\Omega^2$ is the Rayleigh discriminant, $N^2= -\gamma \Omega^2 r D\Theta$ is the square of the centrifugal Br\"{u}nt-V\"{a}iss\"{a}l\"{a} frequency \cite{Kirillov2017}, and $S=Re/Gr$.

The inequality \rf{elsscf1} yields the inviscid LELS criterion
\be{lels}
\frac{N_\Omega^2}{\Omega^2}-\frac{4{\overline{DW}}^2}{1+{\overline{DW}}^2} < 0
\ee
in the isothermal case $\Theta \equiv 0$ \cite{Leibovich}, and the GSF criterion
\be{gsf}
N_{\Omega}^2(1-\gamma\Theta)+Pr N^2 < 0
\ee
in the limit of azimuthal flow $S\rightarrow \infty$ \cite{Kirillov2017,M2013,JT2023}. \textcolor{black}{Thus, \rf{elsscf1} \textit{is a new unified instability criterion} for viscous and thermodiffusive swirling flows (accounting for $Pr$).}

The  envelope in Fig.~\ref{RayleighRBF} closely matches the critical states curve from both linear stability analysis and experiments \cite{Lepiller,Yoshikawa,Guillerm,Kang2015,Kang2023}. The envelope, unlike individual neutral stability curves, has a horizontal asymptote at \(Re = Re_{\infty}\) as $|Gr| \rightarrow \infty$. This explains the seemingly smooth and nearly \(Gr\)-independent stability boundary observed in \cite{Lepiller,Yoshikawa,Guillerm,Kang2015,Kang2023} for $\mu=0$, although this boundary is actually piecewise smooth, with each neutral stability curve touching the common envelope, which flattens at large $|Gr|$ where shear instability dominates. Despite the Rayleigh-Fj\"ortoft shear instability mechanism due to an inflection point in the axial velocity profile \cite{Drazin} leading to axisymmetric perturbations \cite{Bahloul2000,Lepiller2007}, rotation ensures that the critical modes of the BCF are three-dimensional \textcolor{black}{with $k_z\ne 0$ and $m\ne 0$} \cite{Dubrulle2005}.

Figure~\ref{RayleighRBF}, in conjunction with relation \rf{ec}, demonstrates how the envelope selects the critical modes. The lower left branch of the envelope corresponds to the left spiral modes (\(k_z > 0\), \(m > 0\)), while the lower right branch corresponds to the right spiral modes (\(k_z < 0\), \(m > 0\)). This is consistent with findings in \cite{Ali1}.

Computing tangent lines to the envelope in the $(Gr,Re)$-plane at the intersection point \(\rf{re0}\) in the Rayleigh-unstable regime $(Ro < Ro_R)$ yields:
\be{spli}
\frac{Re}{Re_0} = 1 \pm \sqrt{2} \frac{2D\Theta Pr Ro - 3 DW \Theta \gamma k_r^2}{27 Ro k_r^4} Gr.
\ee
This indicates that the temperature gradient is responsible for the existence of two branches that should merge into a single curve for isothermal Rayleigh-unstable flows.

Setting \(\gamma=0\) and \(Ri=0\) in \(\rf{a0}\), we greatly simplify the equation in the isothermal case. This leads to \textit{our second key finding}, which is the simple analytical expression for the envelope  of the neutral stability curves for isothermal swirling flows:
\be{masterisc}
E(Re_z, Re) = \frac{N_{\Omega}^2}{\Omega^2} - \frac{4\overline{DW}^2}{1 + \overline{DW}^2} + \frac{27}{4\Omega^2} \frac{k_r^4}{Re^2} = 0.
\ee
Then, the swirl parameter $S=\frac{Re}{Re_z}$ in \rf{ec}, \rf{a0}, and \rf{masterisc} is redefined via the axial Reynolds number \(Re_z = \frac{W_0 d}{\nu}\), where $W_0$ is the appropriate axial flow velocity.

For example, in the enclosed SCF base flow $W_0=W_1$, the sliding speed of the inner cylinder, the dimensionless axial velocity $W(r)$ is \cite{Ali2,MM2000}:
\ba{escf}
&W(r) = \frac{(2\eta^2\ln \eta - \eta^2 + 1)(r^2 - r_2^2) - (\eta+1)^2\ln(r/r_2)}{((\eta^2 + 1)\ln \eta - \eta^2 + 1)(\eta + 1)(\eta-1)^{-1}},&
\ea
while $V(r)$ is defined by \rf{VT}. The individual neutral stability curves \rf{a0}  and their envelope \rf{masterisc} for SCF are shown in the $(Re_z,Re)$-plane in Fig.~\ref{Figure2}. For Rayleigh-unstable flows ($N_{\Omega}^2<0$ or $Ro < -1$), \rf{masterisc} is a single curve confirming our hypothesis based on  \rf{spli}. The envelope has a maximum $Re=Re_0$ when $Re_z=0$ and a horizontal asymptote $Re=Re_{\infty}$ as $|Re_z|\rightarrow \infty$, where:
 \be{re0i}
     Re_0=\frac{3\sqrt{3}k_r^2}{4\Omega\sqrt{-Ro-1}}, \quad
     Re_\infty=\frac{3\sqrt 3 k_r^2}{4\Omega\sqrt{-Ro}}.
 \ee

As $Ro \rightarrow -1$,  $Re_0 \rightarrow \infty$, and for Rayleigh-stable flows $(-1 < Ro < 0)$, the envelope splits into two curves, each with vertical tangents at $Re_z = \pm Re_z^{\min}$, where
  \be{rezm}
  Re_z^{\min}= \frac{3}{2}\frac{\sqrt{3}k_r^2}{DW}\left(1 + \sqrt{Ro+1}\right)
  \ee
is the critical axial Reynolds number destabilizing Rayleigh-stable isothermal azimuthal flows. Fig.~\ref{Figure2}(c).

The inequality $E(Re_z, Re) < 0$ is the \textit{extension of the LELS criterion \rf{lels} to viscous flows}. In Fig.~\ref{Figure2}(c), criterion \rf{lels} is represented by the oblique solid lines touching the upper parts of the envelope \rf{masterisc} as $Re \rightarrow \infty$. While \rf{lels} does not apply to Rayleigh-unstable flows, its viscous extension defines the stability boundary in this case, as given by the envelope \rf{masterisc}, see Fig.~\ref{Figure2}(a). Setting $\overline{DW} = 0$ in \rf{masterisc} retrieves the stability boundary for isothermal Couette-Taylor flow found in \cite{EY1995}.

The explicit expressions \rf{masterisc} and \rf{re0i} available for isothermal swirling flows allow us to analytically determine the variation of $k_z$ along the stability boundary:
 \be{kze}
     |k_z|(Re)=k_r\frac{\sqrt{2}}{2}\sqrt{\frac{1-Re_{\infty}^2/Re^2}{1-Re_{\infty}^2/Re_0^2}}.
 \ee
Thus, $|k_z|\rightarrow 0$  as $Re \rightarrow Re_{\infty}$ and reaches its terminal value $\frac{\sqrt{2}}{2} k_r$ at $Re = Re_0$ for Rayleigh-unstable flows, Fig.~\ref{Figure2}(b), and $\frac{\sqrt{-2Ro}}{2} k_r$ as $Re \rightarrow \infty$ for Rayleigh-stable flows, Fig.~\ref{Figure2}(d), agreeing with numerical studies \cite{L1964, Ng, MM2000, Ali2}. Additionally, Fig.~\ref{Figure2}(b,d) shows that the spectrum of axial wavenumbers for perturbations differs between Rayleigh-unstable and Rayleigh-stable flows. In Rayleigh-unstable flows, the destabilizing centrifugal force broadens the wavenumber range, whereas in Rayleigh-stable flows, it is limited to lower \( k_z \) values. Long-wavelength perturbations are favored as \( Re \) approaches \( Re_{\infty} \) (and $Re_z \rightarrow \infty$).

\begin{figure}[h]
\centering
\includegraphics[width=.44\textwidth]{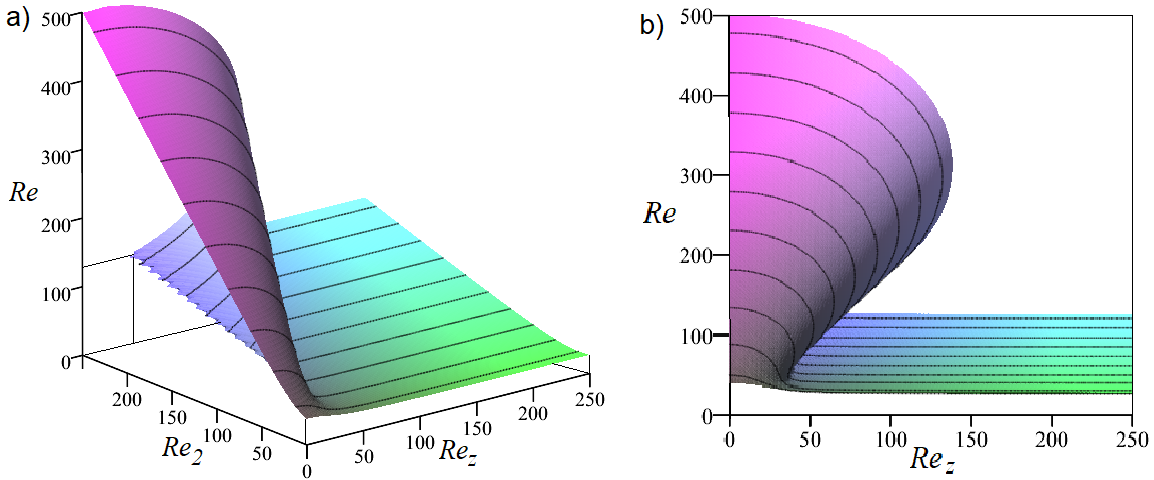}
\caption{For the SCF  \rf{escf} with $k_r = \pi$ and $\eta = 0.5$: (a) The surface of the envelope \rf{masterisc} in the $(Re_z, Re_2, Re)$-space \textcolor{black}{with a pleat and two folds (cf. the cusp catastrophe model \cite{BG1992})} and (b) its projection onto the $(Re_z, Re)$-plane. \textcolor{black}{The color scheme is applied solely to enhance visibility of the surface shape and has no physical significance.} \label{Figure3}}
\end{figure}

The envelope equation \rf{masterisc} is \textit{universal} for all isothermal swirling flows. For swirling flows between differentially rotating cylinders, we express $\mu = \Omega_2 / \Omega_1$ in $\Omega = V / r$ (see \rf{VT}) using the Reynolds numbers of the inner $(Re = \frac{R_1 \Omega_1 d}{\nu})$ and outer ($Re_2 = \frac{R_2 \Omega_2 d}{\nu}$) cylinders as $\mu=\eta\frac{Re_2}{Re}$ \cite{MM2000,Messeguer1}. This reparameterization of Eq.~\rf{masterisc} determines the neutral stability surface in $(Re_z, Re_2, Re)$-space, as shown in Fig.~\ref{Figure3} for the enclosed SCF.
At $Re_z = 0$, the surface yields the neutral stability curve of the Couette-Taylor flow \cite{EY1995}. However, as $Re_z$ increases, the critical $Re$ decreases until it abruptly falls due to a fold first observed numerically for SCF in \cite{MM2000} and then for SPF in \cite{Messeguer1}, despite its different axial shear profile.

\textcolor{black}{This discontinuous transition reminds an analogous phenomenon in the standard magnetorotational instability of magnetohydrodynamics, known as the Velikhov-Chandrasekhar paradox. In this case, the stability boundary of magnetized Couette-Taylor flow drops below the Rayleigh line and approaches the solid-body rotation line when an axial magnetic field is applied, assuming the magnetic Prandtl number—defined as the ratio of viscosity to magnetic diffusivity—differs from one \citep{WB2002, KPS2011, KS2011}.}

\textit{In conclusion}, by employing local geometrical optics stability analysis adapted to visco-diffusive flows, we derived novel explicit instability criteria for isothermal and non-isothermal swirling flows. We found extension of the LELS instability criterion to viscous swirling flows and introduced a new analytical instability criterion for non-isothermal visco-diffusive swirling flows, unifying the LELS and the GSF criteria.

Our advancement stems from an observation overlooked in previous research: the neutral stability curves in these problems possess an envelope, which we have analytically determined using the connection between envelopes and polynomial discriminants. Our analytical results offer a general theory of instabilities across a wide range of swirling flows.

Specifically, our expression \rf{masterisc} for the folded neutral stability surface of isothermal swirling flows elucidates the universal occurrence of the sudden drop in the azimuthal Reynolds number across flows with various shear distributions, highlighting the interplay between shear and centrifugal instability mechanisms.

This work was supported by the French Space Agency (CNES) and the ANR Programme d'Investissements d'Avenir LABEX EMC\(^3\) through the INFEMA project.


\end{document}